\documentclass[aps,prl,superscriptaddress,showpacs,twocolumn]{revtex4} 
\usepackage{amsmath,amssymb,revsymb,epsfig}
\pdfoutput=1 

\usepackage{theorem}

\usepackage{graphicx,latexsym,verbatim,color}

\newlength{\blank}
\newcommand{\ket}[1]{\left | #1 \right \rangle}
\newcommand{\bra}[1]{\left \langle #1 \right |}

\newcommand{\beq}{\begin{equation}}
\newcommand{\eeq}{\end{equation}}

\usepackage{ifthen}
 \usepackage{ifpdf}
 \ifpdf
   \usepackage{epstopdf}
 \else
   
 \fi

\begin{document}

\title{Biased nonlocal quantum games}

\author{Thomas Lawson}
\affiliation{Department of Mathematics, University of Bristol, Bristol BS8 1TW, UK}

\author{Noah Linden}
\affiliation{Department of Mathematics, University of Bristol, Bristol BS8 1TW, UK}

\author{Sandu Popescu}
 \affiliation{H H Wills Physics Laboratory, University of Bristol, Tyndall Ave., Bristol, BS8 1TL, UK}

\date{28th November 2010}

\begin{abstract}
We address the question of when quantum entanglement is a useful resource for
information processing tasks by presenting a new class of nonlocal games that are simple, direct, generalizations of the Clauser Horne Shimony Holt game.
For some ranges of the parameters that specify the games, quantum mechanics offers an advantage, while, surprisingly, for others
quantum mechanics is no more powerful than classical mechanics in performing the nonlocal task.  This sheds new light on the difference
between classical, quantum and super-quantum correlations.
\end{abstract}


\maketitle

A key insight in quantum information theory is that quantum entanglement can be viewed as a resource for information processing tasks.  However, while there has been
considerable progress in how to quantify entanglement (at least bipartite entanglement), we will only fully understand its nature as a resource
when we understand those nonlocal tasks for which it is useful and those for which it is not.  In this Letter we present simple generalizations
of the well-known Clauser-Horne-Shimony-Holt (CHSH) experimental set-up, the set-up which led to the CHSH inequality \cite{Clauser1969}, the most famous Bell-type inequality \cite{Bell1964}.  The new nonlocal tasks we describe are specified by a series of parameters.  As the parameters vary continuously, the tasks undergo a transition from those in which (as in the original CHSH inequality) quantum correlations offer an advantage to those for which quantum nonlocality does not help in performing the nonlocal task.  These tasks shed new light on the nature of quantum nonlocality.  In addition, for almost all values
of the parameters, super-quantum correlations \cite{Popescu1994} perform better than classical or quantum correlations, further emphasizing the subtle nature of the difference between classical, quantum
and super-quantum correlations.

Other interesting scenarios in which quantum correlations are no stronger than classical ones have been described recently \cite{Linden2007,Allcock2009,Almeida2010}.  The particular
interest in the situation that we describe here is its simplicity and close relation to the CHSH game, which it might have been thought was now fully understood; we
point out here that there are intriguing aspects to the CHSH scenario that had not been noticed before.


The CHSH inequality concerns the experimental situation in which a source emits two
particles, one particle sent to a receiver, Alice, and
the other particle sent to a receiver, Bob. Alice and Bob are
located sufficiently far apart that no signal can travel from one
to the other while the experiment is being performed. Alice can
perform one  of two measurements, $A_1$ or $A_2$. Similarly, Bob can
perform $B_1$ or $B_2$. We assume each of these measurements has two possible outcome, $+1$ or $-1$. The experiment is run many times. In each run Alice and Bob choose their measurement at random. The measurements are chosen independently by Alice and Bob; Alice chooses $A_1$ with probability $p$ and $A_2$ with probability $1-p$ while Bob chooses $B_1$ and $B_2$ with probabilities $q$ and $1-q$ respectively.
Let $E(A_iB_j)$ denote the expected value  of the product of the outcomes of the measurements $A_i$ and $B_j$, where $i,j=1,2$.

As shown by Clauser, Horne, Shimony and Holt, any local hidden variables model leads to the inequality
\beq
{1\over4}\bigl(E(A_1B_1) + E(A_1B_2) +E(A_2B_1) - E(A_2B_2)\bigr)
\leq {1\over2}. \label{CHSHineq}
\eeq

Famously, measurements on quantum particles prepared in entangled states may violate the inequality, thus showing that quantum mechanics cannot be modelled by local hidden variables. In other words, no classical system can simulate the quantum correlations because it would involve faster-than-light communication.

The CHSH inequality above can be viewed in two different ways.
In the first approach it doesn't really matter what are the probabilities $p$ and $q$ with which Alice and Bob choose their measurements.  We compute the expectation values $E(A_iB_j)$ by counting in how many cases we obtained $A_iB_j=1$ and in how many cases we obtained $A_iB_j=-1$ and  dividing their difference to the total number of cases $N(i.j)$ in which Alice and Bob happened to choose the pair of measurements $A_i$ and $B_j$, i.e
\beq
E(A_iB_j)={N(A_iB_j=1)-N(A_iB_j=-1)\over N(A_iB_j=1)+N(A_iB_j=-1)}.\label{conditionalexpectation}
\eeq
Thus the CHSH inequality is a relation between {\em conditional} expectation values, $E(A_iB_j)$ of the product of the outcomes of Alice and Bob's measurements, given that the measurements were $ A_i$ and $B_j$.

A second interpretation of \eqref{CHSHineq} is possible in the particular case when $p=q=1/2$.  We can think of the left hand side of this inequality as being the average score in a quantum game, where the average is computed over {\em all} the rounds. In each round Alice and Bob's particles receive a score of +1 or -1. They receive +1 in the following cases:  when Alice and Bob happened to measure $A_1$ and $B_1$, or $A_1$ and $B_2$, or $A_2$ and $B_1$ and the product of their outcomes is +1 and in the case when they  measured $A_2$ and $B_2$ and the product of their outcomes is -1. In all other cases, they receive -1.  Indeed, the factor 1/4 represents the probability of any particular pair of measurements $A_i$ and $B_j$ and the expectation values taken with the corresponding sign are the average scores, given that the corresponding measurements were performed. The inequality \eqref{CHSHineq} represents the maximum average score that Alice and Bob's particles can obtain if they are classical. Quantum particles in entangled states, subjected to appropriate quantum measurements, violate this inequality, i.e. they can perform better in this game than any classical systems.

The game, as presented above, is important for many communication and computation tasks. Adopting the point of view of a computer scientist, we
 can think of the experiment as an input-output problem. Alice and Bob have each a binary variable, $x$ and $y$ respectively. They feed their variables to a system (that consists of their particles and measuring devices) and they receive a binary output, $a$ and $b$ respectively. If we identify the input $x=0$ with the instruction \lq\lq measure $A_1$\rq\rq\ and $x=1$ as \lq\lq measure $A_2$\rq\rq\   and if we identify the result +1 of Alice's with the output $a=0$ and the result -1 with $a=1$  and do similar identifications for Bob, then the above game is mapped to the following: Alice and Bob's particles (as a team) win if they  output $a$ and $b$ such that
\beq
a\oplus b=xy
\eeq
where $\oplus$ denotes addition modulo 2.

Inequality \eqref{CHSHineq} can be easily re-written as an inequality for the average probability of success of this game,
\beq
\sum_{x,y=0}^1 {1\over4} P(a\oplus b=xy|xy) \leq {3\over4}
\eeq

We come now to the main question of this paper: what happens if the probabilities $p$ and $q$ with which Alice and Bob choose their measurements are not equal to one half? What is the average score of the game and can quantum particles always perform better than classical ones?

Of course, if our only goal is to verify whether quantum mechanics is nonlocal or not, the probabilities $p$ and $q$ are irrelevant because, as explained above,  all we have to do is to compute the conditional averages and use them in \eqref{CHSHineq}.  But in communication and computation tasks every single round counts, so the overall probability of success is the relevant quantity.

Thus we now consider how well Alice and Bob's particles can perform  in a game whose score is
\begin{eqnarray}
\sum_{x,y=0}^1 P(x,y) P(a\oplus b=xy|xy),
\end{eqnarray}
where $P(x,y)$ is the probability that the input pair is $(x,y)$.  Alice and Bob's particles are assumed to know the distribution $P(x,y)$
which is specified at the start of the game.  Thus it may be to the particles' advantage to modify their joint strategy to take into
account that some inputs may occur more frequently and so \lq\lq winning\rq\rq\ for those inputs is likely to give a higher overall
score.

Let us consider first the distribution discussed in the introduction $P(x,y)=P(x)P(y)$ where $P(x=0)=p,P(y=0)=q$.
In order to compute the best classical and quantum scores for this game it will be convenient
to use the (entirely equivalent) formalism using operators $A_i$ and $B_i$.  The expression
we wish to maximise is
\begin{eqnarray}\label{CHSH-pq}
& & CHSH[p,q]=pqE(A_1B_1) + p(1-q)E(A_1B_2)\\
& &\qquad +q(1-p)E(A_2B_1) - (1-p)(1-q)E(A_2B_2).\nonumber
\end{eqnarray}

We first treat the case that $p,q\geq 1/2$.

General arguments show that the maximum classical value occurs at an extremal strategy; for example
the strategy where the  values of all the observables are $+1$ is optimal and achieves the value
\begin{eqnarray}
1-2(1-p)(1-q).\label{CHSHclassicalmax}
\end{eqnarray}

We now consider quantum strategies.  We first find a bound on
quantum strategies independent of the dimension of Hilbert space.  Using standard
techniques (Schwartz's inequality) one can show that
\begin{eqnarray}
& &CHSH_{Q}[p,q]
\leq   p\sqrt{q^2+(1-q)^2 + q(1-q)\alpha} \nonumber\\
& &\qquad\qquad+(1-p)\sqrt{q^2+(1-q)^2 - q(1-q)\alpha},\label{Tsirelson-pq}
\end{eqnarray}
where
\begin{eqnarray}
\alpha= \bra\psi I\otimes(B_1B_2 + B_2B_1)\ket\psi,
\end{eqnarray}
where $B_i$ are Hermitian operators on Bob's Hilbert space satisfying $B^2_i=I$, and $\ket\psi$ is an arbitrary pure state
in Alice and  Bob's Hilbert space.

Finding the maximum of \eqref{Tsirelson-pq} for general $p$ and $q$ is a little more subtle than it is
for the usual case $p=q=1/2$.  The issue is that the maximum value of
$\alpha$ is 2, so that, depending on the values of $p$ and $q$, the maximum value of \eqref{Tsirelson-pq}
may occur inside the region $0\leq \alpha < 2$ or at the boundary $\alpha=2$.  In fact the maximum occurs at
\begin{eqnarray}
\alpha_{max}= \min\left(2, \frac{(q^2+(1-q)^2)(p^2-(1-p)^2)}{q(1-q)(p^2+(1-p)^2)}\right).
\end{eqnarray}
We note that
\begin{eqnarray}
\frac{(q^2+(1-q)^2)(p^2-(1-p)^2)}{q(1-q)(p^2+(1-p)^2)}&=& 2\nonumber\\
\Rightarrow pq&=&\frac{1}{2}.
\end{eqnarray}

Thus we have two regions of $[p,q]$ space to consider.

\bigskip

{\bf Region 1: $1\geq p \geq (2q)^{-1}\geq 1/2$}

Here the maximum value of \eqref{Tsirelson-pq} occurs at $\alpha=2$.  This may be achieved
by taking $B_1=B_2$.  For this value of $\alpha$,
\begin{eqnarray}
CHSH_{Q}[p,q]
\leq 1- 2(1-p)(1-q).
\end{eqnarray}

In other words, in this region, the bound we have found for quantum particles
is the same as that achievable by classical particles. Thus, in this region, quantum
mechanics provides no benefit in playing this nonlocal game.

\bigskip

{\bf Region 2: $1\geq (2q)^{-1}> p \geq 1/2$}

In this region the bound is achieved at
\begin{eqnarray}
\alpha=\frac{(q^2+(1-q)^2)(p^2-(1-p)^2)}{q(1-q)(p^2+(1-p)^2)}.
\end{eqnarray}
Substituting this value of $x$ into \eqref{Tsirelson-pq}, we find the bound for quantum
systems to be:
\begin{eqnarray}
CHSH_{Q}[p,q]
\leq \sqrt{2}\sqrt{q^2+(1-q)^2}\sqrt{p^2+(1-p)^2}.
\end{eqnarray}

We note that this is greater than the classical bound \eqref{CHSHclassicalmax} in
region 2.  All that remains to be shown is that there exists a quantum strategy achieving this
bound.  It may be checked that the following choices suffice:
\begin{eqnarray}
A_1&=&\frac{X(q+(1-q)\cos\beta)+Z(1-q)\sin\beta }{\sqrt{(q+(1-q)\cos\beta)^2+(1-q)^2\sin^2\beta}},
\nonumber\\
A_2&=&\frac{X(q-(1-q)\cos\beta)-Z(1-q)\sin\beta}{\sqrt{(q-(1-q)\cos\beta)^2+(1-q)^2\sin^2\beta}},
\nonumber\\
B_1&=&X,\nonumber\\
B_2&=&X\cos\beta  + Z\sin\beta,\nonumber\\
 \ket\psi&=&{1\over \sqrt 2} \left(\ket 0\ket 0 +\ket 1\ket
1\right),
\end{eqnarray}
where $X$ and $Z$ are the standard Pauli operators, and
\begin{eqnarray}
\cos\beta=
\frac{1}{2}\frac{(q^2+(1-q)^2)(p^2-(1-p)^2)}{q(1-q)(p^2+(1-p)^2)}.
\end{eqnarray}
The parts of $p,q$ space where one or both of $p$ and $q$ less than one half can be treated
similarly; as expected the situation is highly symmetric as illustrated in Fig. 1.

There has been much interest recently in considering super-quantum correlations; correlations which are stronger than quantum mechanics, but are nonetheless
non-signalling \cite{Barrett2005}.  It is known that non-signalling correlations
(\lq\lq nonlocal boxes\rq\rq) can
achieve the algebraic maximum value, 1, for the usual CHSH game \eqref{CHSHineq}.  The same
is easily seen for \eqref{CHSH-pq}: extremal nonlocal boxes can also achieve the algebraic
maximum, 1, for our generalized games \eqref{CHSH-pq}.  Thus for all $p,q\neq 0,1$ the games are nonlocal, in the sense
that generalised non-signalling correlations can give advantage over local strategies.

\begin{figure}[tb]
\centering
\includegraphics[scale=0.8]{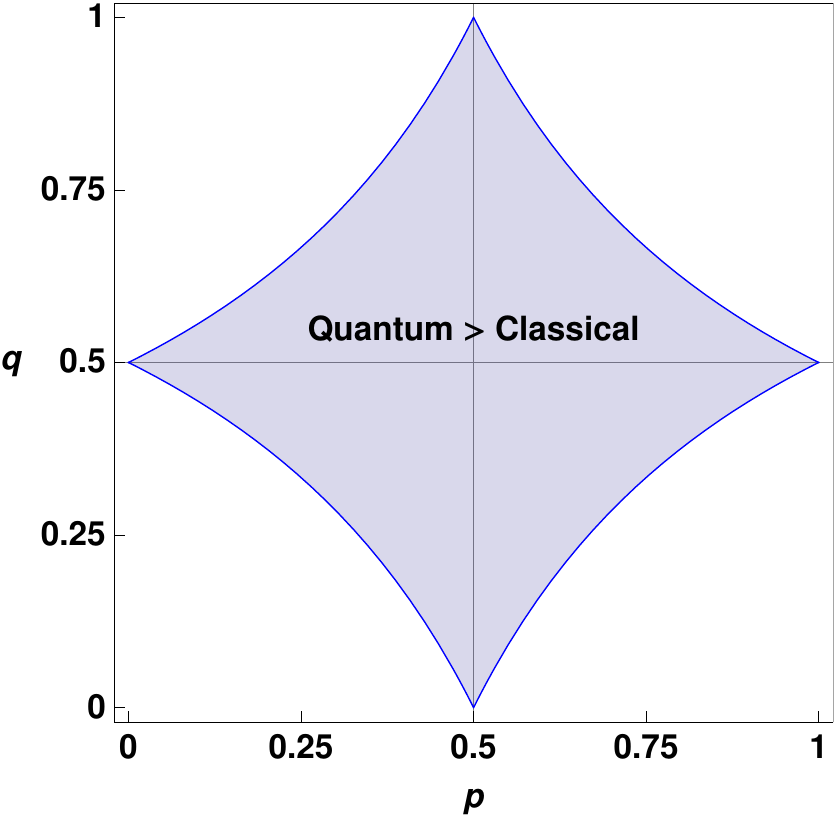}
\caption{Quantum strategies can perform better than classical strategies in the shaded region of $p,q$ space, but
not outside;  the games are nonlocal for all $p,q\neq 1$ }\label{figure:pq-plot}
\end{figure}

We now present a number of generalizations of these ideas.  Firstly one could
imagine that rather than Alice and Bob independently controlling the probability
of their measurement choice, it could be that there is a \lq\lq game controller\rq\rq  who
chooses  $(x,y)$ with a probability distribution $P(x,y)$ that need not be a product
distribution.  Thus we consider maximizing the expression
\begin{eqnarray}
& &P_{00}P(a\oplus b=0|00) + P_{01}P(a\oplus b=0|01)\\
 & &\quad +P_{10}P(a\oplus b=0|10) + P_{11}P(a\oplus b=1|11),\nonumber
\end{eqnarray}
where we have written $P(x=i,y=j)=P_{ij}$.  Without loss of generality, we consider the case $P_{00}\ge P_{01}\ge P_{10}\ge P_{11}$.
Very similar (but slightly more involved) calculations to those above show that the condition that
quantum strategies are no better than classical ones is
\begin{eqnarray}
{1\over P_{00}} + {1\over P_{01}} + {1\over P_{10}} - {1\over P_{11}} \leq 0.
\end{eqnarray}

\begin{figure}[tb]
\centering
\includegraphics[scale=0.9]{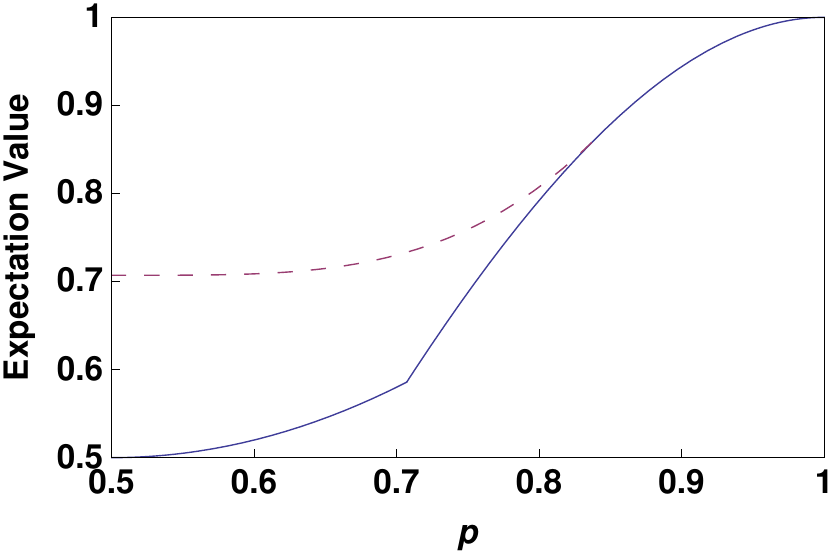}
\caption{Numerical evaluation of the quantum (shown in dashed red) and classical optimal strategies for the Svetlichny inequality, $S_3[p]$. There is no quantum advantage above $p\simeq 0.8406$}\label{figure:Svetlichny qu and cl graph}
\end{figure}

A further generalization is to more parties.  A particularly interesting family of
inequalities is the family \cite{Collins2002} of $n$-party Svetlichny \cite{Svetlichny1986}
inequalities.  We now generalize these to the situation that each party has a biased
probability for the two measurements to be made.
We thus consider $n$ parties; each party $i$ measures one of two observables, denoted $C_i$ and  $C^\prime_i$;
the observables have outcomes $\pm 1$.
For each party $i$ the measurement $C_i$ is made with probability $p$ and $C^\prime_i$ is made with probability $1-p$.
The known inequalities \cite{Collins2002} correspond to $p=1/2$.
The expression  $S_n[p]$ for the $n$ parties that we are interested in is most easily defined via the relations:
\begin{eqnarray}
& &M_2[p]=2\Big(p^2 C_1 C_2 + p(1-p) C_1 C^\prime_2 \\
& &\qquad+ (1-p)p C^\prime_1 C_2-(1-p)^2 C^\prime_1 C^\prime_2\Big),\nonumber\\
& &M_{n+1}[p]= p M_n[p]C_{n+1}+pM_n'[p]C_{n+1}\\
& &\qquad+(1-p)M_n[p] C^\prime_{n+1}-(1-p) M_n'[p] C^\prime_{n+1},\nonumber
\end{eqnarray}
where $M_n'[p]$ is obtained from $M_n[p]$ by exchanging all the primed and non-primed $C$'s.  Finally we define
\begin{eqnarray}
S_n[p]&=& \left\{
\begin{array}{lcl} M_n[p]
&,&\mbox{$n$ even}\\ {1\over 2}(M_n[p]+M_n'[p]) &,&\mbox{$n$ odd}
\end{array}
\right.\,.
\end{eqnarray}
We are interested in the maximum expectation value of $S_n[p]$ for classical and quantum strategies.  For the quantum
maximum, in principle one has to maximize over states in arbitrary dimension. Fortunately, our inequalities are of the
form where the results of \cite{Wolf2001} may be used to provide a considerable simplification: it suffices to use qubits
for each particle, the particles being in a generalized GHZ state \cite{Greenberger1990} and one needs only to maximise
over $n+1$ angles $\phi_0,\phi_1,...,\phi_n$. The operators may be taken to be
\begin{eqnarray}
C_k&=&\cos(\frac{\phi_0}{n})X + \sin(\frac{\phi_0}{n})Y\\
C^\prime_k&=&\cos(\phi_k+\frac{\phi_0}{n})X + \sin(\phi_k+\frac{\phi_0}{n})Y,
\end{eqnarray}
where  $X$ and $Y$ are the Pauli operators.
We used numerical optimization over the angles $\phi_k$ to find the maximum quantum value.  The maximum
classical value was also computed numerically for each $p$ by exhaustive calculation of the value for each extremal strategy.  These numerically computed optima for $S_3[p]$ are
shown in Fig. \ref{figure:Svetlichny qu and cl graph}.

\begin{figure}[tb]
\centering
\includegraphics[scale=0.8]{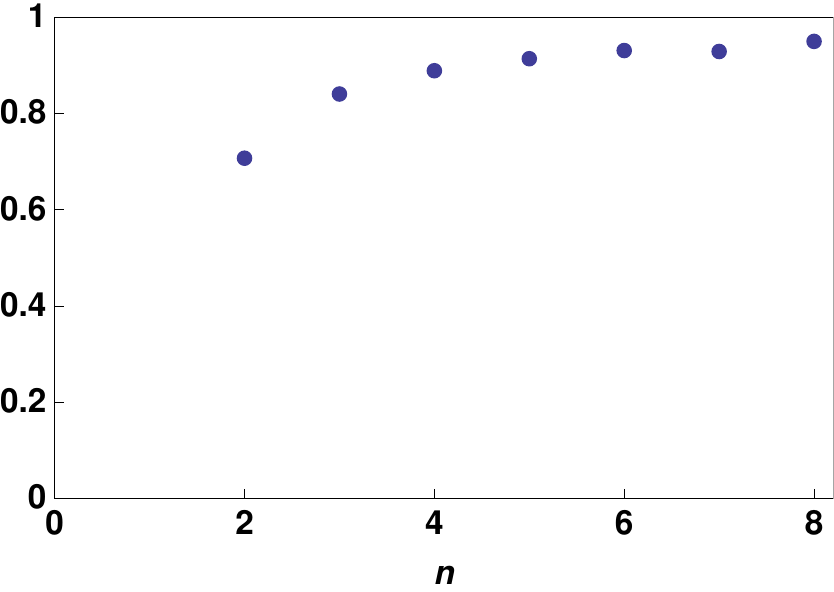}
\caption{Numerical evaluation of the value of $p$ above which there is no quantum mechanical advantage over local hidden variables theories against system size $n$ for the Svetlichny inequalities.}\label{figure:p up to 10}
\end{figure}

It is also interesting to look at the minimum value
of $p$ for each number of parties $n$ for which quantum strategies give no benefit over classical ones.  For small $n$, this is plotted
in Fig. \ref{figure:p up to 10}.

We end by making some observations, firstly concerning \eqref{CHSH-pq}. While we have proven that there
is a region in which quantum mechanics is more powerful than classical mechanics
in playing these nonlocal games, it would be valuable to have some intuition
as to why this should be so.  For example, why should quantum mechanics not help
play the game when, say, $p=q=3/4$?   A further interesting case is the game when
$[p,q]=[1-\epsilon,1/2]$, for small $\epsilon$.  In this case the game is barely nonlocal:
Alice is almost certain to measure $A_1$.  Nonetheless quantum mechanics
provides a benefit in this case.

Our proof, that there is a quantum state which achieves the bound in the region where
quantum mechanics is stronger than classical mechanics, used a maximally entangled state.
Since the game is not maximally nonlocal, it would be interesting to know whether we can
achieve the quantum bound with less entanglement.

Finally, it would be very interesting to
have some intuition as to why the minimum value of $p$ for which quantum strategies
do not out-perform classical ones  increases as $n$ increases, and also understand the
limit $n\rightarrow \infty$.

\medskip
\begin{acknowledgments}
 We thank the  UK EPSRC for support through the QIP-IRC.
\end{acknowledgments}


\begin{thebibliography}{10}

\bibitem{Clauser1969} J.F.~Clauser, M.A.~Horne, A.~Shimony and R.A.~Holt, Phys. Rev. Lett. {\bf 23}, 880 (1969).
\bibitem{Bell1964} J.~Bell, Physics {\bf 1}, 1995 (1964).
\bibitem{Popescu1994} S.~Popescu and D.~Rohrlich, Found. Phys. {\bf 24}, 379 (1984).
\bibitem{Linden2007} N.~Linden, A.~Short, S.~Popescu and A.~Winter, Phys. Rev. Lett. {\bf 99}, 180502 (2007).
\bibitem{Allcock2009}J.~Allcock, H.~Buhrmann and N.~Linden, Phys. Rev. A {\bf 80}, 032105 (2009).
\bibitem{Almeida2010} M.L.~Almeida, J.~Bancal, N.~Brunner, A.~Acin, N.~Gisin and S.~Pironio, Phys. Rev. Lett. {\bf 104}, 230404 (2010)
\bibitem{Collins2002} D.~Collins, N.~Gisin, S.~Popescu, D.~Roberts and V.~Scarani, Phys. Rev. Lett. {\bf 88}, 170405 (2002); M.~Seevinck and G.~Svetlichny, Phys. Rev. Lett. {\bf 89}, 060401 (2002).
\bibitem{Svetlichny1986} G.~Svetlichny, Phys. Rev. D {\bf 35}, 3066 (1987).
\bibitem{Wolf2001} R.F.~Werner and M.M.~Wolf, Phys. Rev. A, {\bf 64}, 032112 (2001).
\bibitem{Greenberger1990} D.M.~Greenberger, M.~Horne, A.~Shimony and A.~Zeilinger, Am. J. Phys. {\bf 58}, 1131 (1990)
\bibitem{Barrett2005} J.~Barrett, N.~Linden, S.~Massar, S.~Pironio, S.~Popescu and D.~Roberts, Phys. Rev. A {\bf 71}, 022101 (2005).



\end{thebibliography}

\end{document}